\documentclass[%
 reprint,
superscriptaddress,
 amsmath,amssymb,
pre,
]{revtex4-2}

\usepackage{graphicx}
\usepackage{dcolumn}
\usepackage{bm}
\usepackage{color}
\usepackage[colorlinks=true,allcolors=blue,breaklinks=true]{hyperref}
\usepackage{physics}
\usepackage{tabularx}

\begin{document}

\preprint{APS/123-QED}

\title{Designing the pressure-dependent shear modulus using tessellated granular metamaterials}

\author{Jerry Zhang}
\thanks{These authors contributed equally.}
\affiliation{
 Department of Mechanical Engineering \& Materials Science, Yale University, New Haven, Connecticut 06520, USA
}

\author{Dong Wang}
\thanks{These authors contributed equally.}
\affiliation{
 Department of Mechanical Engineering \& Materials Science, Yale University, New Haven, Connecticut 06520, USA
}

\author{Weiwei Jin}
\affiliation{
 Department of Mechanical Engineering \& Materials Science, Yale University, New Haven, Connecticut 06520, USA
}

\author{Annie Xia}
\affiliation{
 Department of Mechanical Engineering \& Materials Science, Yale University, New Haven, Connecticut 06520, USA
}

\author{Nidhi Pashine}
\affiliation{
 Department of Mechanical Engineering \& Materials Science, Yale University, New Haven, Connecticut 06520, USA
}

\author{Rebecca Kramer-Bottiglio}
\affiliation{
 Department of Mechanical Engineering \& Materials Science, Yale University, New Haven, Connecticut 06520, USA
}

\author{Mark D. Shattuck}
\affiliation{
 Benjamin Levich Institute and Physics Department, The City College of New York, New York, New York 10031, USA
}

\author{Corey S. O'Hern}
\email{corey.ohern@yale.edu}
\affiliation{
 Department of Mechanical Engineering \& Materials Science, Yale University, New Haven, Connecticut 06520, USA
}
\affiliation{
 Department of Physics, Yale University, New Haven, Connecticut 06520, USA
}
\affiliation{
 Department of Applied Physics, Yale University, New Haven, Connecticut 06520, USA
}
\affiliation{
 Integrated Graduate Program in Physical and Engineering Biology, Yale University, New Haven, Connecticut 06520, USA
}

\date{\today}

\begin{abstract}
Jammed packings of granular materials display complex mechanical response. For example, the ensemble-averaged shear modulus $\left\langle G \right\rangle$ increases as a power-law in pressure $p$ for static packings of soft spherical particles that can rearrange during compression. We seek to design granular materials with shear moduli that can either increase {\it or} decrease with pressure without particle rearrangements even in the large-system limit. To do this, we construct {\it tessellated} granular metamaterials by joining multiple particle-filled cells together.  We focus on cells that contain a small number of bidisperse disks in two dimensions. We first study the mechanical properties of individual disk-filled cells with three types of boundaries: periodic boundary conditions (PBC), fixed-length walls (FXW), and flexible walls (FLW). Hypostatic jammed packings are found for cells with FLW, but not in cells with PBC and FXW, and they are stabilized by quartic modes of the dynamical matrix. The shear modulus of a single cell depends linearly on $p$. We find that the slope of the shear modulus with pressure, $\lambda_c < 0$ for all packings in single cells with PBC where the number of particles per cell $N \ge 6$. In contrast, single cells with FXW and FLW can possess $\lambda_c > 0$, as well as $\lambda_c < 0$, for $N \le 16$. We show that we can force the mechanical properties of multi-cell granular metamaterials to possess those of single cells by constraining the endpoints of the outer walls and enforcing an affine shear response. These studies demonstrate that tessellated granular metamaterials provide a novel platform for the design of soft materials with specified mechanical properties.
\end{abstract}

\maketitle

\section{Introduction}
\label{intro}

Granular materials represent an interesting class of physical systems that are composed of individual macroscopic particles that interact via dissipative, contact forces~\cite{jaeger96rmp}. As a result of the dissipative particle interactions, granular materials come to rest in the absence of external driving, such as applied shear or vibration. Because of this, they frequently occur in amorphous states lacking long-range positional order. Further, granular systems can undergo a jamming transition, where they develop nonzero bulk {\it and} shear moduli when they are compressed to large packing fractions~\cite{jamming:OHernPRE2003, liu10review, vanHecke10review}.

There have been numerous computational~\cite{edwards89, jamming:OHernPRE2003, silbert05prl, olsson07prl, henkes07prl, song08nature, aste08prl, henkes09pre, mailman09prl, schreck10sm, manning11prl, goodrich14pre, goodrich16pnas, patinet16prl, martiniani17natphys, jamming:VanderWerfPRE2018, jamming:VanderWerfPRL2020} and experimental~\cite{doney04science, corwin05nature, dauchot05prl, majmudar07prl, keys07natphys, clusel09nature, bi11nature, dijksman11prl, puckett13prl, miskin13natmat, lin16prl} studies of the structural and mechanical properties of jammed granular materials. In particular, it has been shown that the shear modulus depends sensitively on structural disorder, non-affine particle motion, the number of contacts and anisotropy of the interparticle contact network~\cite{goodrich14natphys, maloney06pre, ellenbroek09epl, ellenbroek06prl, vanHecke10review, zaccone11prb, jamming:WangPRE2021, zhang21pre, dagoisbohy12prl, goodrich12prl, goodrich14pre, goodrich16pnas, mizuno16prl, zaccone14jap, schlegel16scirep, zhang23pre:localG}.  For example, in jammed packings of frictionless spherical particles with purely repulsive linear spring interactions, we have shown that the shear modulus $G$ decreases linearly with pressure $p$ along ``geometrical families,'' where the network of interparticle contacts does not change during isotropic compression~\cite{jamming:VanderWerfPRL2020}. If a particle rearrangement occurs during the compression at $p = p^*$, \textit{e.g.} through the addition of an interparticle contact, $G$ jumps discontinuously at $p^*$ and the linear relation between $G$ and $p$ no longer holds. Also, when a particle rearrangement occurs, it is difficult to predict the new interparticle contact network and the mechanical properties of the jammed packing are no longer reversible. The range of pressure $\Delta p$ over which the contact network does not change decreases with increasing system size, $\Delta p \sim N^{-1}$, where $N$ is the number of particles in the system. Thus, in the large-$N$ limit, granular packings undergo frequent irreversible particle rearrangements to new jammed packings after each $\Delta p$ increment. During compression, each new contact network typically possesses an increased number of contacts, and thus the shear modulus increases with pressure. In fact, studies have shown that the ensemble-averaged shear modulus scales as $\langle G\rangle \sim p^{0.5}$ in the large $pN^2$ limit for jammed packings of spherical particles with purely repulsive linear spring interactions~\cite{goodrich14pre}.

In this article, we design granular metamaterials for which the shear modulus can either decrease or increase with increasing pressure with no particle rearrangements. In linear elastic solids, the shear modulus does not depend on the pressure. In conventional atomic and molecular solids, both the bulk and shear moduli increase with pressure at large pressures~\cite{thompson78glassmoduli, antao08prl, zeng18materials}. Similarly, in conventional granular materials, the shear modulus increases with pressure due to the formation of new contacts during compression, but it is history-dependent. Our design of granular metamaterials will leverage the recent findings that particle rearrangements in granular packings with small $N$ are rare and the shear modulus depends linearly on pressure in the absence of rearrangements~\cite{jamming:VanderWerfPRL2020}. Preventing particle rearrangements ensures reversibility of the packing’s mechanical properties and improves our ability to predict them.  We will first consider systems in two dimensions (2D), but these concepts can easily be extended to three dimensions (3D). For example, we have shown that the pressure-dependent shear modulus for jammed packings of spherical particles is qualitatively the same in 2D and 3D~\cite{jamming:VanderWerfPRL2020}.

\begin{figure}
    \centering
    \includegraphics[width=\linewidth]{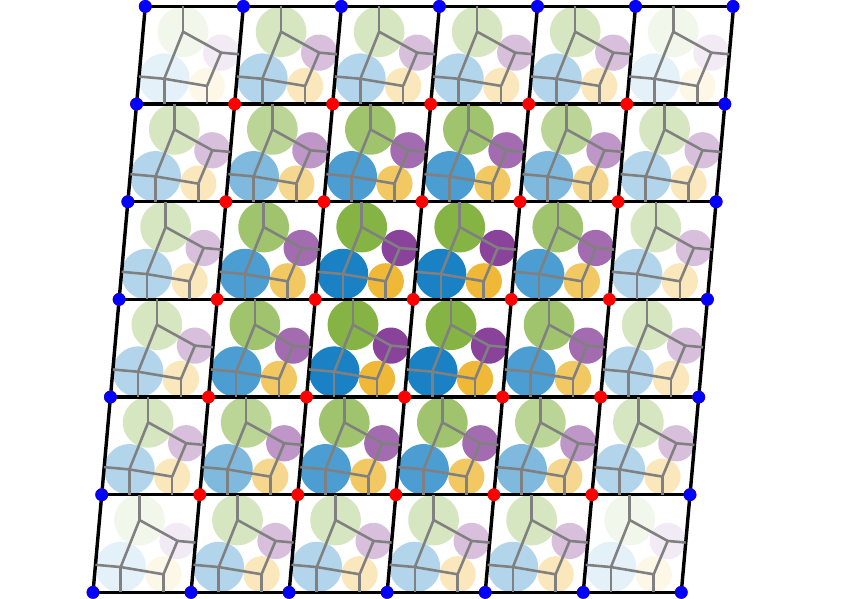}
\caption{Illustration of a {\it tessellated} granular metamaterial, made up of $36$ individual cells. Each cell contains the same jammed bidisperse packing of $N=4$ disks that are confined by four freely jointed, flexible walls. The interior cells share all four walls and the edge cells share three walls. To generate the collection of disk-filled cells, we first create a disk packing within a single cell, connect multiple copies of this disk-filled cell, fix the outer blue vertices, and then allow the disks and interior red vertices to relax during energy minimization. The variation in the disk shading between different cells indicates the types of cells based on their adjacent cells. For example, cells on the right edge of the tessellation only have three adjacent cells. The cell type is determined by its distance from the four outer walls and four corner vertices.}
\label{fig:trussexample}
\end{figure}

We envision tessellated granular metamaterials that are made up of many individual cells that each contain a small number of grains, i.e. $N <16$, and are bounded by four freely jointed elastic walls. The disks within each cell are jammed with typically an isostatic number of contacts. See Fig.~\ref{fig:trussexample}. The mechanical response of each cell is highly anisotropic, i.e., its shear modulus depends on the angle $\theta$ of the applied shear relative to the orientation of the confining walls. We find that the shear modulus of each cell obeys $G_c = G_{c0} + \lambda_c p$, where $G_c=G_{c0}$ at $p=0$, and we determine the sign and magnitude of $\lambda$ as a function $\theta$, $N$, and the ratio of the particle and wall stiffnesses. We vary the size of the tessellated granular metamaterials by adding multiple copies of individual cells together, e.g. by generating an $n \times n$ array of cells that share the confining walls. We identify the regimes where the shear modulus of the full system is similar to that for the individual cells.  In particular, we find that large tessellated granular metamaterials can possess shear moduli that {\it decrease} with increasing pressure, i.e. the opposite behavior compared to conventional granular materials, and that these materials retain the anisotropy of the individual cells.

The remainder of the article is organized as follows. In Sec.~\ref{sec:methods}, we describe the computational methods, including the particle-particle, particle-wall, and wall-wall potential energies, the protocols for generating disk-filled single cells (henceforth referred to as ``cells'') and collections of multiple cells, and the methods for calculating the pressure, shear stress, and shear modulus of these structures. In Sec.~\ref{sec:results}, we present the results on how the boundary conditions, individual disk packing configuration, and the ratio of the particle to wall stiffness affect the relation between the shear modulus and pressure in single cells, as well as coupled systems composed of $\mathcal{N}_c = n^2$ cells. In Sec.~\ref{sec:conclusions}, we provide conclusions and discuss promising directions of future research, such as the mechanical response of tessellated granular metamaterials in three dimensions. We also include three Appendices. In Appendix~\ref{sec:isostaticity}, we show that Maxwell-like counting arguments can be used to determine the minimum number of particle-particle and particle-wall contacts in jammed disk packings within single cells with fixed-length and flexible walls, and explain the occurrence of ``quartic modes'' in cells with flexible walls. In Appendix~\ref{sec:C_mat_rot}, we determine analytical expressions for the dependence of the components of the stiffness matrix on the angle of the applied simple shear strain for jammed disk packings in single cells. In Appendix~\ref{sec:twoparticle}, we verify that the pressure-dependence of the single-cell shear modulus is related to the second derivative of the packing fraction at jamming onset $\phi_J$ with respect to shear strain $\gamma$ for an example cell with fixed-length walls. 

\section{Methods}
\label{sec:methods}

\begin{figure}
    \centering
    \includegraphics[width=\linewidth]{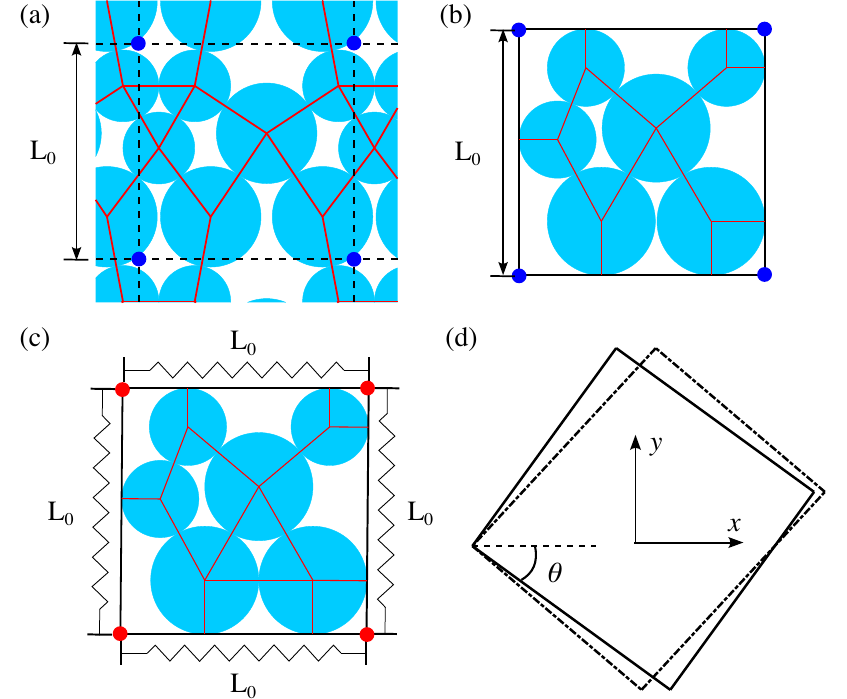}
\caption{Illustration of cells that contain $N=6$ bidisperse disks with three different boundary conditions: (a) periodic boundary conditions (PBC) in a square cell with side length $L_0$, (b) a cell with four straight walls of fixed length $L_0$ (FXW), and (c) a cell with four flexible walls (FLW) such that adjacent vertices are connected by linear springs with preferred length $L_0$. To generate jammed disk packings within each cell, we successively compress the system, fixing the blue vertices after the compression, and then allowing the disks and red vertices to relax. (d) Illustration of the application of simple shear strain $\gamma = 0.2$ to an originally square cell (solid line) at angle $\theta$ relative to the $x$-axis, which generates the parallelogram-shaped cell indicated by the dashed-dotted line.}
    \label{fig:shearexample}
\end{figure}

We study individual cells containing jammed packings of $N$ bidisperse soft, frictionless disks: $N/2$ small and $N/2$ large disks with diameter ratio $\sigma_l/\sigma_s=1.4$. A diameter ratio of $\sigma_l/\sigma_s=1.4$ gives rise to disordered jammed disk packings~\cite{liu10review, zhang15jcp}. We consider three types of boundary conditions for the cells as illustrated in Fig.~\ref{fig:shearexample}: (a) periodic boundary conditions (PBC) in square cells with side length $L_0$, (b) cells with four straight walls of fixed length $L_0$ (FXW), and (c) cells with four flexible walls (FLW) such that adjacent vertices are connected by linear springs with preferred length $L_0$. For boundary condition (c), the connected walls are freely jointed such that the angle between them can change without energy cost. 

We model the tessellated granular metamaterials using linear spring interaction potentials (either purely repulsive or double-sided), which are commonly used in discrete element method simulations of granular materials~\cite{liu10review}. Frictionless disks that interact via pairwise, purely repulsive linear spring forces are placed within each cell. The corresponding interparticle potential energy is given by
\begin{equation}
    \label{eq:energy_pp}
    U^{pp}(r_{jk}^{pp}) = \frac{\epsilon_{pp}}{2} \left(1 - \frac{r^{pp}_{jk}}{\sigma_{jk}} \right)^2 \Theta \left(1 - \frac{r^{pp}_{jk}}{\sigma_{jk}}\right),
\end{equation}
where $\epsilon_{pp}$ gives the strength of the repulsive interactions, $r^{pp}_{jk}$ is the distance between the centers of disks $j$ and $k$, $\sigma_{jk}$ is the sum of the radii of disks $j$ and $k$, and $\Theta(\cdot)$ is the Heaviside step function. The repulsive force on disk $j$ from $k$ is ${\vec f}^{pp}_{jk} = -(dU^{pp}/dr^{pp}_{jk}) {\hat r}^{pp}_{jk}$, where ${\hat r}^{pp}_{jk}$ is a unit vector pointing from the center of disk $k$ to the center of disk $j$. Previous studies have shown that the soft particle model in Eq.~\ref{eq:energy_pp} generates the same disk packings at jamming onset at those for rigid disks~\cite{gao09pre}.

For PBC boundary condition (a), there are only interparticle interactions. For boundary conditions (b) (FXW) and (c) (FLW), we also consider repulsive interactions between the disks and walls using the purely repulsive linear spring potential energy,
\begin{equation}
    \label{eq:energy_pb}
    U^{pb}(r^{pb}_{ji}) = \frac{\epsilon_{pb}}{2} \left(1 - \frac{r^{pb}_{ji}}{R_j} \right)^2 \Theta \left(1 - \frac{r^{pb}_{ji}}{R_j}\right),
\end{equation}
where $\epsilon_{pb}$ is the strength of the repulsive interactions between the disks and walls, $r^{pb}_{ji}$ is the shortest distance between the center of disk $j$ and the $i$th wall, and $R_j$ is the radius of disk $j$. The repulsive force on disk $j$ from the $i$th wall is ${\vec f}^{pb}_{ji} = -(dU^{pb}/dr^{pb}_{ji}) {\hat r}^{pb}_{ji}$, where ${\hat r}^{pb}_{ji}$ is the unit vector pointing to the center of disk $j$ and perpendicular to the $i$th wall. 

For FLW boundary conditions, we consider interactions between the wall endpoints using the double-sided linear spring potential energy,
\begin{equation}
    \label{eq:energy_bb}
    U^{bb}(r^{bb}_{i, i+1}) = \frac{\epsilon_{bb}}{2} \left(1 - \frac{r^{bb}_{i, i+1}}{L_0} \right)^2,
\end{equation}
where $\epsilon_{bb}$ is the characteristic energy scale of the linear spring potential,  $r^{bb}_{i, i+1}$ is the distance between endpoints $i$ and $i + 1$, and $L_0$ is equilibrium length for the $i$th wall. The force on endpoint $i$ from endpoint $i+1$ in the $i$th wall is ${\vec f}^{bb}_{i} = -(dU^{bb}_i/dr^{bb}_{i,i+1}) {\hat r}^{bb}_{i,i+1}$, where ${\hat r}^{bb}_{i,i+1}$ is the unit vector pointing from endpoint $i+1$ to $i$. 

We calculate the stress tensor $\Sigma_{\alpha \beta}$ (with $\alpha$, $\beta =x$, $y$) of the tessellated granular metamaterials using the virial expression~\cite{allen17compsimliquids}. For cells with PBC, the total potential energy is $U = \sum^N_{j<k} U^{pp}(r^{pp}_{jk})$ and stresses are generated only from interparticle forces. The total stress tensor is thus $\Sigma_{\alpha \beta} = \Sigma^{pp}_{\alpha \beta}$, where
\begin{equation}
    \label{eq:stress_pp}
    \Sigma^{pp}_{\alpha \beta} = \frac{1}{A}\sum^N_{j>k}f^{pp}_{jk\alpha}r^{pp}_{jk\beta},
\end{equation}
$A$ is the area of the cell, $f^{pp}_{jk\alpha}$ is the $\alpha$-component of the force on disk $j$ from $k$, and $r^{pp}_{jk\beta}$ is the $\beta$-component of the separation vector from the center of disk $k$ to the center of disk $j$. 

For cells with physical walls as shown in Fig.~\ref{fig:shearexample} (b) and (c), the forces between the walls and particles also contribute to the stress tensor. For cells with FXW boundary conditions, the total potential energy is $U = \sum^N_{j<k} U^{pp}(r^{pp}_{jk}) + \sum^N_{j=1} \sum^4_{i=1} U^{pb}(r^{pb}_{ji})$. In this case, the total stress tensor is $\Sigma_{\alpha \beta} = \Sigma^{pp}_{\alpha \beta} + \Sigma^{pb}_{\alpha \beta}$, where
\begin{equation}
    \label{eq:stress_pb}
    \Sigma^{pb}_{\alpha \beta} = \frac{1}{A} \sum^4_{i=1} \sum^N_{j=1}f^{pb}_{ji\alpha}r^{pb}_{ji\beta},
\end{equation}
$f^{pb}_{ji\alpha}$ is the $\alpha$-component of the force on disk $j$ from the $i$th wall of the cell, and $r^{pb}_{ji\beta}$ is the $\beta$-component of the separation vector from the contact point between wall $i$ and disk $j$ to the center of disk $j$. 

For cells with FLW boundary conditions, in addition to the interparticle and particle-wall interactions, the walls store potential energy. Thus, the total potential energy is $U = \sum^N_{j<k} U^{pp}(r^{pp}_{jk}) + \sum^N_{j=1} \sum^4_{i=1} U^{pb}(r^{pb}_{ji}) + \sum^4_{i=1} U^{bb}(r^{bb}_{i, i+1})$. The total stress tensor is $\Sigma_{\alpha \beta} = \Sigma^{pp}_{\alpha \beta} + \Sigma^{pb}_{\alpha \beta} + \Sigma^{bb}_{\alpha \beta}$ for cells with flexible walls, where
\begin{equation}
    \label{eq:stress_bb}
    \Sigma^{bb}_{\alpha \beta} = \frac{1}{A} \sum^4_{i=1}f^{bb}_{i\alpha}r^{bb}_{i\beta},
\end{equation}
$f^{bb}_{i\alpha}$ is the $\alpha$-component of the spring force from wall $i$, and $r^{bb}_{i\beta}$ is the $\beta$-component of the vector with the same length as wall $i$ pointing in the same direction as ${\vec f}^{bb}_{i}$. The pressure of the cell is $p = (\Sigma_{xx} + \Sigma_{yy}) / 2$ and the shear stress is $\Sigma=-\Sigma_{xy}$. We use $\epsilon_{pp}/\sigma_s^2$ for the units of stress and shear modulus and $\epsilon_{pp}$ for units of energy. 

To generate jammed disk packings within a single cell, we first place $N$ disks randomly in the cell at a dilute packing fraction, $\phi < 10^{-3}$. We then apply an affine isotropic compressive strain to the disk positions and decrease the length of the walls by $\Delta L$ to achieve a small packing fraction increment, $\Delta \phi/\phi = 2\Delta L/L_0 = 2 \times 10^{-3}$, followed by potential energy minimization using the fast inertia relaxation engine (FIRE) algorithm \cite{fire}. During energy minimization, the disk positions change, while the endpoints of the fixed-length walls for the boundary conditions depicted in Fig.~\ref{fig:shearexample} (a) and (b) are held fixed. However, the endpoints of the flexible walls in Fig.~\ref{fig:shearexample} (c) are allowed to move during energy minimization. After energy minimization, we calculate the pressure $p$ of the cell. If $p$ is less than the target pressure $p_t = p_t^j = 10^{-7}$, we again compress the system by $\Delta \phi/\phi$ and perform energy minimization. If $p > p_t^j$, we return to the previous disk and wall configuration, compress the system by $\Delta \phi/2$, and perform energy minimization. We repeat this process until the cell pressure satisfies $|p - p_t^j|/ p_t^j < 10^{-4}$. Our results do not change if we choose smaller values of $p_t^j$. For all three boundary conditions in Fig.~\ref{fig:shearexample} (a)-(c), we generate $10^4$ disk packings at jamming onset.

To investigate the mechanical response of the cells as a function of pressure, we apply isotropic compression to the cells at jamming onset to achieve a range of pressure values $p_t^c$ that are logarithmically spaced between $10^{-7}$ to $10^{-2}$. To ensure that the shape of the cells does not significantly deviate from the cell shape at jamming onset, we fix the endpoints of the walls for all three types of boundary conditions when generating the cells with pressures above jamming onset $p_t^j$. Since we are using energy minimization to generate the packings, nearly all packings that we obtain at various $p_t^c$ are mechanically stable.

To calculate the shear modulus of a single cell $G_c$ at an angle $\theta$ relative to the $x$-axis, we first rotate the cell clockwise by $\theta$, as shown in Fig.~\ref{fig:shearexample} (d). Determining $G_c(\theta)$ allows us to assess the anisotropy of the mechanical response of single cells. We then apply successive small steps of simple shear strain $\Delta\gamma = 5 \times 10^{-9}$ (where $x$ is the shear direction and $y$ is the shear gradient direction) to the disks and walls with each strain step followed by potential energy minimization. Note that after the applied simple shear strain, the walls remain fixed during energy minimization for all three boundary conditions. We obtain the shear modulus for a single cell by calculating $G_c = d\Sigma_c/d\gamma$, where $\Sigma_c$ is the shear stress of a single cell. Over the range of shear strain used to measure $G_c$, $\Sigma_{xy}(\gamma)$ is linear. In practice, linear fits of $\Sigma_{xy}(\gamma)$ yield R-squared values that satisfy $1 - R^2 < 10^{-11}$.

We build large-scale tessellated granular metamaterials by joining multiple copies of a given cell with flexible walls at jamming onset, e.g. the collection of $36$ coupled cells in Fig.~\ref{fig:trussexample}. After joining the cells, we perform potential energy minimization with the outermost (blue) wall endpoints held fixed, while the internal (red) endpoints, as well as the disk positions, are allowed to relax. Disks within a given cell only interact with other disks and the walls of that cell. Interior wall endpoints have four connections to other walls, while exterior wall endpoints have either two or three connections to other walls. We generate tessellated granular metamaterials at jamming onset with $p=p_t^j=10^{-7}$, compress the systems to achieve pressures that are logarithmically spaced between $p_t^j$ and $10^{-2}$, and measure the shear modulus at each pressure. The shear modulus $G$ of the collection of cells is calculated in the same way as that for a single cell. In particular, we first rotate the aggregate by $\theta$ clockwise, and then we apply small successive steps of simple shear strain, $\Delta\gamma = 5 \times 10^{-9}$, with each step followed by energy minimization, where the outer vertices are held fixed and the inner vertices, as well as the disks, are allowed to relax. The total shear stress $\Sigma$ of the tessellated granular metamaterial is the sum of $\Sigma^{pp}$ and $\Sigma^{pb}$ for all cells and the unique contributions to $\Sigma^{bb}$ for all of the cell walls. The shear modulus of the tessellated granular metamaterial is given by $G = d\Sigma/d\gamma$. When compressing the tessellated metamaterials to increase $p$, or applying simple shear to measure $G$, we impose a specified displacement field on the particle and wall positions prior to energy minimization. After imposing the displacement field, we allow the particles and only the internal wall vertices to relax their positions under potential energy minimization.

After we apply each simple shear strain step followed by energy minimization to tessellated granular metamaterials, which can in principle give rise to non-affine displacements, we calculate the displacement field $\mathcal{F}_{pq}$ of all cell wall endpoints to characterize the non-affine displacements of each cell.  We find the strain field that minimizes the total non-affine displacement of all endpoints for a given cell and simple shear strain step~\cite{falk98pre}:
\begin{equation}
    \label{eq:d2min}
    \mathcal{F}_{pq} = \sum_{s=x,y} X_{ps} \left( Y^{-1} \right)_{sq},
\end{equation}
where
\begin{equation}
    X_{ps} = \sum^4_{i=1} r_{ci,p} r^0_{ci,s},
\end{equation}
and
\begin{equation}
    Y_{sq} = \sum^4_{i=1} r^0_{ci,s} r^0_{ci,q}.
\end{equation}
Here, $r^0_{ci,s}$ and $r_{ci,s}$ are the $s$th component of the separation vector from the center of mass of a given cell to its $i$th endpoint before and after the applied simple shear strain, respectively. We subtract the applied simple shear strain $\gamma$ from $\mathcal{F}_{xy}$ to determine the non-affine displacement field.

\section{Results}
\label{sec:results}

\begin{figure*}
    \centering
    \includegraphics[width=\linewidth]{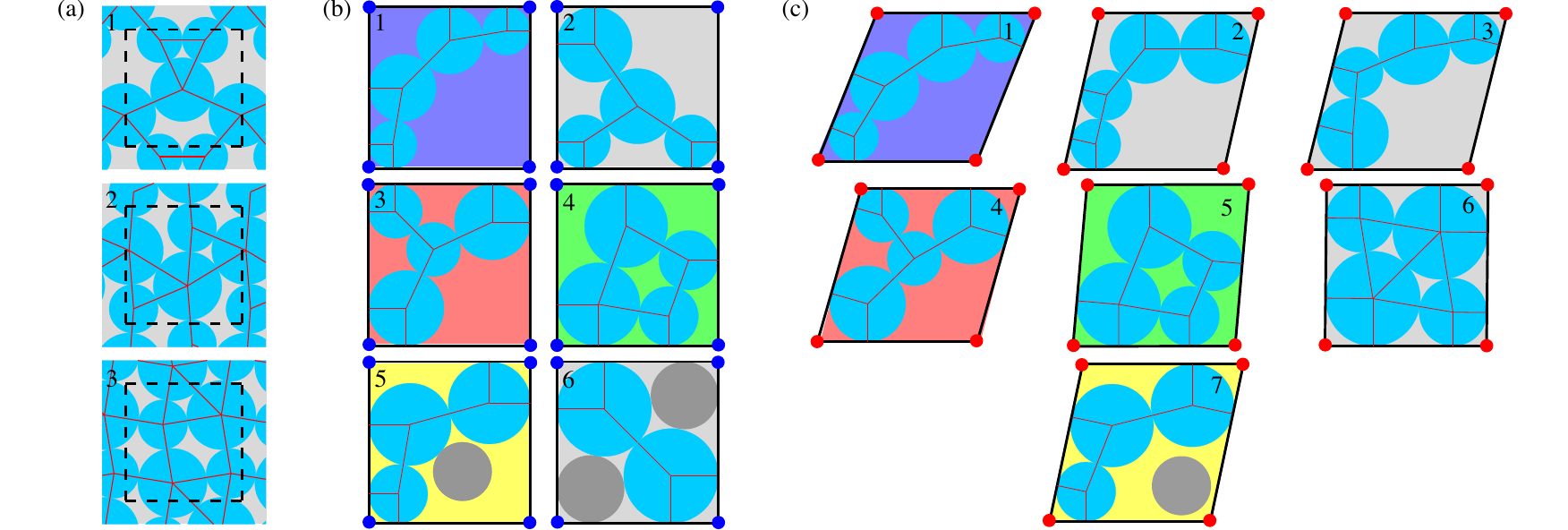}
\caption{Distinct $N=4$ bidisperse cells at \textit{jamming onset} for (a) periodic boundary conditions (PBC, dashed lines), (b) fixed-length physical walls (FXW), and (c) flexible physical walls (FLW). The solid red lines indicate interparticle contacts. The non-gray background colors indicate shared interparticle contact networks in panels (b) and (c). The disks shaded in dark gray are `rattler' particles with no interparticle contacts.}
    \label{fig:fourparticle_config}
\end{figure*}

In this section, we describe the results for the mechanical response of single cells, as well as large collections of cells.  
In Sec.~\ref{subsec:jam_fourparticle}, we enumerate all of the distinct $N = 4$ bidisperse disk packings in single cells at jamming onset for all three boundary conditions. We determine whether the shear modulus for single cells $G_c$ increases or decreases with pressure over the full range of $\theta$ in Sec.~\ref{subsec:gp_single}. We find that $G_c$ for cells with PBC nearly always decreases with pressure (for all shear angles), while $G_c$ can either decrease or increase with pressure for single cells with (both FXW and FLW) physical walls. We further show that the slope of the shear modulus versus pressure $\lambda_c = dG_c/dp$ for single cells can be tuned by varying the particle-wall interaction energy $\epsilon_{pb}$ and wall stiffness $\epsilon_{bb}$. Finally, in Sec.~\ref{subsec:gp_cgm}, we emphasize that the sign and magnitude of $\lambda_c$ for a single cell can be maintained even in a large collection of cells since the assembly prevents particle rearrangements. We then show that the mechanical response of large collections of cells can deviate from the single-cell behavior when we allow the outer cell walls to relax and change their positions during energy minimization.

\subsection{Single cells with $N = 4$}
\label{subsec:jam_fourparticle}

\subsubsection{Jammed configurations}
\label{subsubsec:configuration}

We first illustrate the different types of jammed bidisperse disk packings that occur in single cells with PBC and physical walls. 
In Fig.~\ref{fig:fourparticle_config}, we show all possible jammed cells with $N = 4$. We find $3$ distinct jammed packings for single cells with PBC~\cite{gao06pre}, $6$ distinct packings for cells with FXW~\cite{ashwin12pre}, and $7$ distinct packings for cells with FLW. For $N=2$, there is only one distinct jammed cell with the disks arranged along the diagonal of the cell.  

For jammed cells with FLW, the shape of the boundary is not typically a square, as shown in Fig.~\ref{fig:fourparticle_config} (c), since the energy function for the walls does not include a bending energy term. Despite this, we show that several of the jammed configurations in the cells with FXW and FLW share the same interparticle contact networks, e.g. configuration $1$ in Fig.~\ref{fig:fourparticle_config} (b) and (c).

For $N = 4$, we find that rattler particles occur in jammed cells with FXW and FLW. See configurations $5$ and $6$ in Fig.~\ref{fig:fourparticle_config} (b) and configuration $7$ in Fig.~\ref{fig:fourparticle_config} (c). Rattler disks also occur for cells with PBC and physical walls for $N>4$. Since our focus is on jammed packings that do not undergo particle rearrangements during simple shear and isotropic compression, we will not include calculations of the mechanical response for cells with rattler disks.

\subsubsection{Contact Number}
\label{subsubsec:contact}

The boundary conditions of the cells affect the number of interparticle contacts that are required at the onset of jamming. The numbers of degrees of freedom for the cells are the following: PBC, $N_{d} = 2N' - 1$, FXW, $N_{d} = 2N' + 1$, and FLW, $N_{d} = 2N' + 2$, where $N' = N - N_r$ and $N_r$ is the number of rattler disks. (See Appendix~\ref{sec:isostaticity}.) For mechanically stable disk packings~\cite{maxwell64, pellegrino86stability}, the number of contacts must satisfy $N_c \ge N_d$. Packings with $N_c = N_d$ are isostatic; packings with $N_c < N_d$ are hypostatic; and packings with $N_c > N_d$ are hyperstatic. For $N = 4$, in PBC, we find that the jammed bidisperse disk packings are either isostatic ($N_c = N_d$, $N' = 4$, $N_c = 7$ for configuration $1$) or hyperstatic ($N_c > N_d$, $N' = 4$, $N_c = 8$ for configuration $2$, and $N' = 4$, $N_c = 9$ for configuration $3$). In cells with FXW, all of the packings are isostatic ($N' = 4$, $N_c = 9$ for configurations $1$-$4$, $N' = 3$, $N_c = 7$ for configuration $5$, and $N' = 2$, $N_c = 5$ for configuration $6$). In the cells with FLW, most of the jammed bidisperse disk packings are {\it hypostatic} ($N_c < N_d$, $N' = 4$, $N_c = 9$ for configurations $1$-$5$ and $N' = 3$, $N_c = 7$ for configuration $7$). As discussed in Sec.~\ref{subsubsec:quartic}, quartic modes are required to stabilize hypostatic jammed packings. In contrast, configuration $6$ in Fig.~\ref{fig:fourparticle_config} (c) is hyperstatic ($N' = 4$, $N_c = 13$).

\subsubsection{Quartic Modes}
\label{subsubsec:quartic}

Hypostatic jammed packings have only been reported for packings of non-spherical particles~\cite{mailman09prl, jamming:VanderWerfPRE2018} and particles with shape and size degrees of freedom~\cite{brito18pnas, treado21prm, wang21sm}. Our results indicate that jammed packings of spherical particles can also be hypostatic in cells with FLW. We have shown in previous studies that jammed hypostatic packings are stabilized by quartic modes~\cite{mailman09prl, jamming:VanderWerfPRE2018, treado21prm}, which do not occur in isostatic and hyperstatic packings. Indeed, we find that hypostatic cells at jamming onset possess $N_d-N_c$ quartic modes. (See Appendix~\ref{sec:isostaticity}.) For $N > 4$, we also find that jammed disk packings are isostatic in cells with PBC, either isostatic or hyperstatic for cells with FXW, and either isostatic, hyperstatic or hypostatic for cells with FLW. At large $N$ ($N > 16$), we find jammed disk packings are typically isostatic in all types of boundary conditions studied. (See Appendix~\ref{sec:isostaticity}.)

\subsection{Shear modulus versus pressure for a single cell}
\label{subsec:gp_single}

\subsubsection{Slope of shear modulus with pressure and anisotropy}
\label{subsubsec:single_slope}

\begin{figure*}
    \centering
\includegraphics[width=\linewidth]{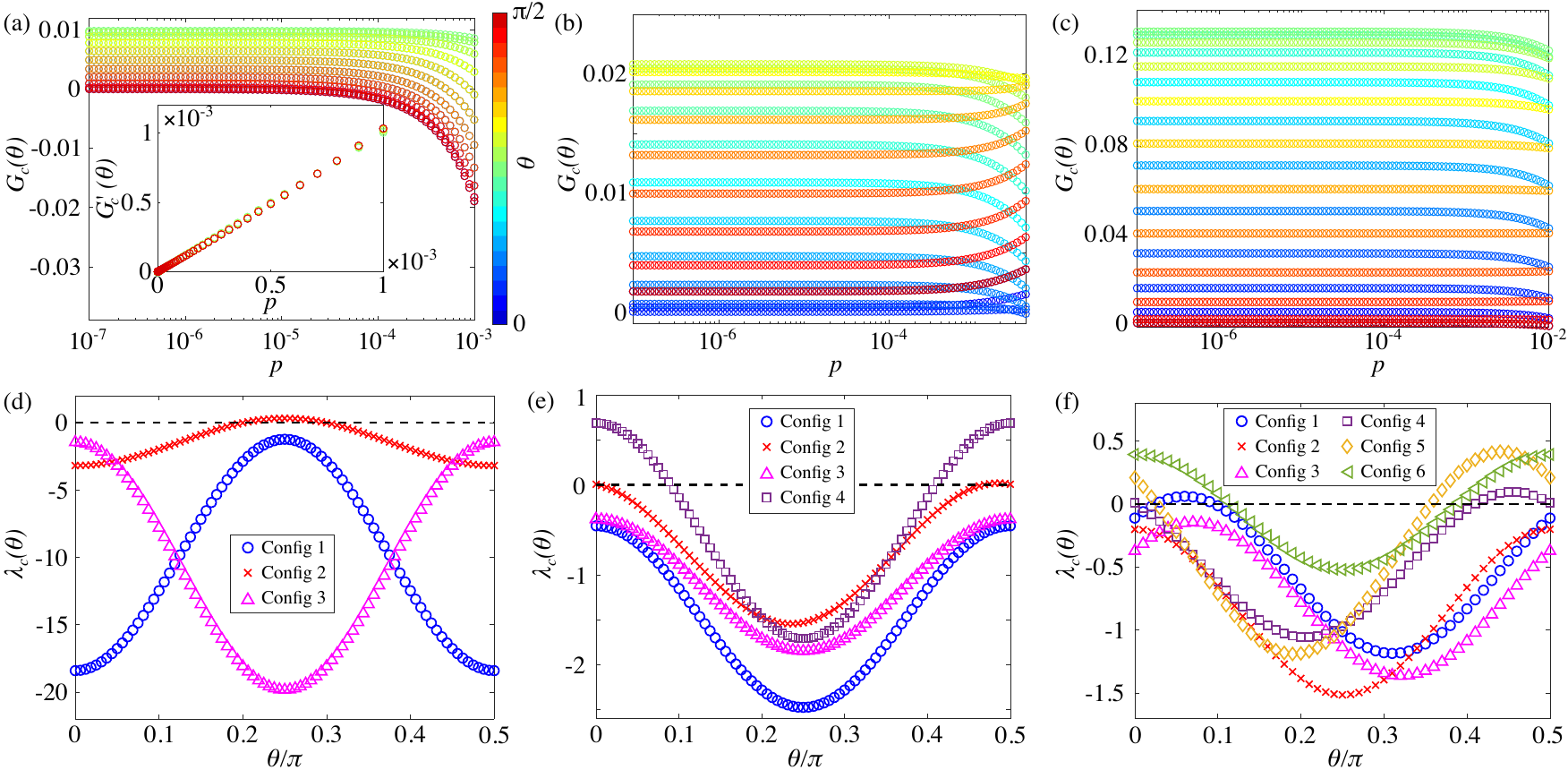}
\caption{(a)-(c) Shear modulus $G_c(\theta)$ versus pressure $p$ for a single cell in configuration $1$ (PBC) in Fig.~\ref{fig:fourparticle_config} (a), configuration $4$ (FXW) in Fig.~\ref{fig:fourparticle_config} (b), and configuration $5$ (FLW) in Fig.~\ref{fig:fourparticle_config} (c). The color of each curve indicates the angle $\theta$ of the applied shear strain, which varies from $0$ (blue) to $\pi/2$ (red). The inset gives $G'_c(\theta)=(G_c(\theta)-G_{c0}(\theta))/\lambda_c(\theta)$ for the data in the main panel of (a). (d)-(f) The slope of $G_c(\theta)$ versus $p$, $\lambda_c(\theta)$, plotted as a function of $\theta/\pi$ for the configurations in Fig.~\ref{fig:fourparticle_config} (except those with rattler disks). The different colors and symbols correspond to different configurations of the cells with (d) PBC, (e) FXW, and (f) FLW (blue circles: configuration $1$; red crosses: $2$; purple upward triangles: $3$; magenta squares: $4$; yellow diamonds: $5$; and green left triangles: $6$). The horizontal dashed lines indicate $\lambda_c = 0$.}
    \label{fig:fourparticle_sm}
\end{figure*}

In Fig.~\ref{fig:fourparticle_sm} (a)-(c), we show the shear modulus $G_c(\theta)$ of single cells as a function of pressure $p$ over the full range of shear angles $\theta$ for cells with PBC, FXW, and FLW, respectively. In contrast to the behavior for large-$N$ systems, we find that the disks do not rearrange and $G_c(\theta)$ varies continuously with $p$ over more than four orders of magnitude. For cells with PBC, $G_c(\theta)$ typically decreases with $p$ as shown in Fig.~\ref{fig:fourparticle_sm} (a). In contrast, for cells with FXW (Fig.~\ref{fig:fourparticle_sm} (b)) and FLW (Fig.~\ref{fig:fourparticle_sm} (c)), $G_c(\theta)$ can either decrease or increase with $p$, depending on the value of $\theta$.

As we showed previously for jammed packings of spherical particles with PBC, we find quite generally that $G_c(\theta)$ varies linearly with $p$~\cite{jamming:VanderWerfPRL2020, zhang21pre}, $G_c(\theta)=G_{c0}(\theta) +\lambda_{c}(\theta)p$, for cells with PBC and physical walls in the absence of particle rearrangements (see the inset to Fig.~\ref{fig:fourparticle_sm} (a)). $G_{c0}(\theta)$ gives the single-cell shear modulus in the zero-pressure limit and $\lambda_c(\theta) =dG_c(\theta)/dp$ gives the slope~\cite{jamming:VanderWerfPRL2020}. In Fig.~\ref{fig:fourparticle_sm} (d)-(f), we plot $\lambda_c(\theta)$ as a function of $\theta$ for all $N = 4$ cells without rattlers. We show that $\lambda_c(\theta) = \lambda_{c,a}\sin[4(\theta-\theta_0)] + \lambda_{c,dc}$ varies sinusoidally with period $\pi/2$, where $\lambda_{c,a}$ is the amplitude, $\theta_0$ is the phase shift, and $\lambda_{c,dc}$ is the mean value of $\lambda_c(\theta)$~\cite{goodrich12prl}. (Previous studies have shown that the shear modulus of jammed packings of spherical particles is sinusoidal with period $\pi/2$~\cite{goodrich14pre, zhang23pre:localG}.) $\lambda_c(\theta) <0$ for nearly all $\theta$ values and cells for PBC, except for configuration $2$ (Fig.~\ref{fig:fourparticle_config} (a)) in the range $0.2 \lesssim \theta/\pi \lesssim 0.3$ (Fig.~\ref{fig:fourparticle_sm} (d)). For cells with FXW and FLW, we observe similar sinusoidal behavior for $\lambda_c(\theta)$, but there are large $\theta$ ranges where $\lambda_c(\theta)>0$. Our results showing that $\lambda_{c,a} \sim \lambda_{c,dc}$ emphasize that cells at small $N$ are highly anisotropic. For cells with FLW, we do not find a correlation between $\lambda_c(\theta) > 0$ and the occurrence of quartic modes as discussed in  Sec.~\ref{subsubsec:quartic}.

\begin{figure}
    \centering
\includegraphics[width=\linewidth]{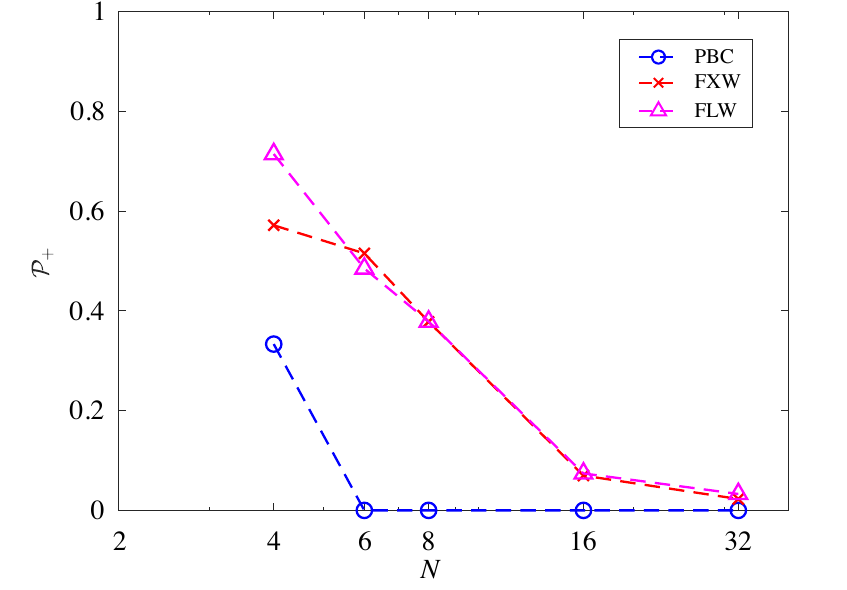}
\caption{Probability ${\cal P}_+$ that cells possess $\lambda_c(\theta)>0$ for any nonzero range of $\theta$, as a function of system size $N$ for PBC (blue circles), FXW (red crosses), and FLW (pink triangles).}
    \label{fig:posprob}
\end{figure}

\begin{figure*}
    \centering
\includegraphics[width=\linewidth]{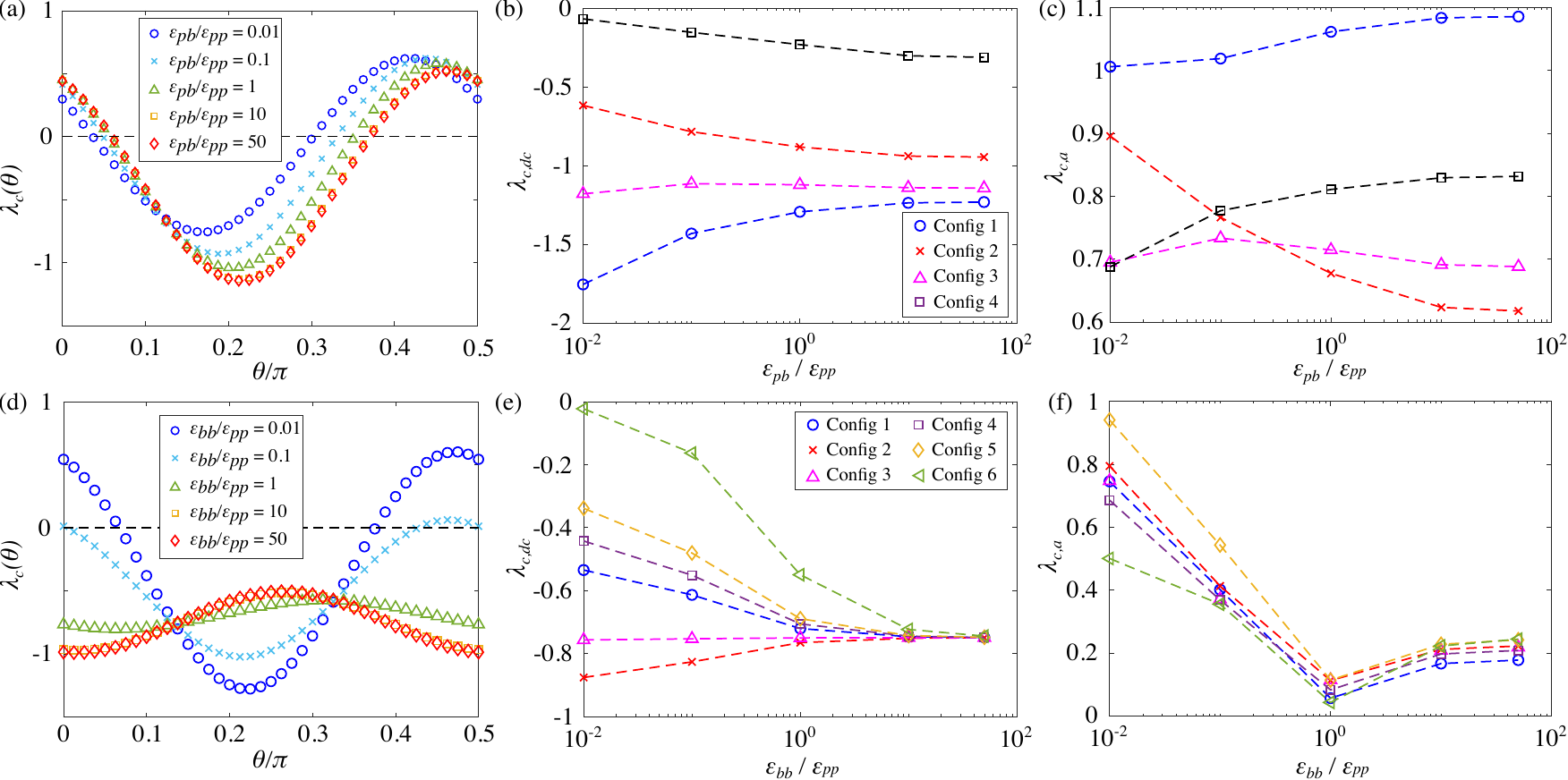}
\caption{(a) The slope of the shear modulus $\lambda_c(\theta)$ versus pressure for a single cell with FXW (configuration $4$ in Fig.~\ref{fig:fourparticle_config} (b)) versus the angle $\theta$ of the applied shear strain for several values of the particle-wall interaction energy normalized by the strength of the interparticle repulsive energy, $\epsilon_{pb}/\epsilon_{pp}$ indicated by the different colors and symbols. (b) Mean value $\lambda_{c,dc}$ and (c) amplitude $\lambda_{c,a}$ of the slope of the shear modulus as a function $\epsilon_{pb}/\epsilon_{pp}$ for the four single cells without rattlers and FXW shown in Fig.~\ref{fig:fourparticle_config} (b). The symbols and colors in (b) and (c) indicate the specific configurations in Fig.~\ref{fig:fourparticle_config} (b). (d)-(f) Similar data to that in panels (a)-(c) except for single cells with FLW. Data for configuration $5$ in Fig.~\ref{fig:fourparticle_config} (c) is shown in panel (d). The symbols and colors in (e) and (f) indicate the specific configurations in Fig.~\ref{fig:fourparticle_config} (c). The dashed horizontal lines in (a) and (d) indicate $\lambda_c=0$.}
    \label{fig:boundarystiffness}
\end{figure*}

Isotropic, linearly elastic materials in 2D possess only two elastic moduli, \textit{i.e.} the bulk and shear moduli. In this study, we focus on the response of the system to simple shear strain. However, small, jammed disk packings at low pressures are anisotropic~\cite{mizuno16prl, zhang23pre:localG}, and thus they possess more than two elastic constants that are inter-related and depend on $\theta$. In Appendix~\ref{sec:C_mat_rot}, we derive the $\theta$-dependence for all six stiffness matrix elements $\hat{C}$ and show that the elements of $\hat{C}$ are related to each other under rotations. In contrast, for isotropic, linearly elastic materials, $\hat{C}_{33} \equiv G$ is independent of $\theta$. Therefore, in future studies we will consider all elements of $\hat{C}$ to fully understand the pressure- and angle-dependent elastic moduli for anisotropic materials. In addition, our previous studies have shown that the sign of $\lambda_c(\theta=0)$ is determined by the second derivative of the packing fraction at jamming onset with respect to $\gamma$ in jammed packings of spherical particles in PBC~\cite{jamming:VanderWerfPRL2020, zhang21pre}. In Appendix~\ref{sec:twoparticle}, we show that this relation is still true at any $\theta$ in cells with fixed-length walls. In particular, this relationship expresses $\hat{C}_{33}$ in terms of derivatives of the packing fraction, $U$, and $p$ with respect to shear strain for any given $\theta$ and determines which terms in this expression depend most strongly on $p$. Similar analyses will be carried out for the other elements of $\hat{C}$ in future studies to gain a complete understanding of the pressure-dependent elastic moduli of small jammed disk packings.

\subsubsection{Probability of $\lambda_c(\theta) > 0$}
\label{subsubsec:p_plus}

As $N$ increases, the probability ${\cal P}_+$ to obtain a cell with $\lambda_c(\theta) >0$ decreases rapidly for PBC. As shown in Fig.~\ref{fig:posprob}, we do not find $\lambda_c(\theta) >0$ for cells with $N \ge 6$ for PBC. For cells with physical walls, $\mathcal{P}_{+}$ also decreases with increasing $N$, but not as rapidly as that for cells with PBC. These results emphasize that if one wants to tune the pressure dependence of $G_c(\theta)$ (i.e. between $\lambda_c(\theta) >0$ and $\lambda_c(\theta) <0$), one should employ cells with small $N$.

\subsubsection{Effects of wall and particle stiffnesses on $\lambda_c(\theta)$}
\label{subsubsec:wall_stiffness}

We next investigate the dependence of $\lambda_c(\theta)$ on the particle-wall stiffness $\epsilon_{pb}/\epsilon_{pp}$ and wall stiffness $\epsilon_{bb}/\epsilon_{pp}$ relative to the strength of the repulsive interparticle interactions in cells with FXW and FLW. In Fig.~\ref{fig:boundarystiffness} (a), we show that for configuration $4$ depicted in Fig.~\ref{fig:fourparticle_config} (b), $\lambda_{c,a}$, $\lambda_{c,dc}$, and $\theta_0$ undergo only small variations when $\epsilon_{pb}/\epsilon_{pp}$ changes by nearly two orders of magnitude. $\lambda_{c,a}$ and $\lambda_{c,dc}$ converge for $\epsilon_{pb}/\epsilon_{pp} \gtrsim 10$ for all $N=4$ cells with FXW (Fig.~\ref{fig:boundarystiffness} (b) and (c)).  For configuration $4$, we find that $\lambda_c(\theta)>0$ for a finite range of $\theta$ in the large $\epsilon_{pb}/\epsilon_{bb}$ limit.  We can also fix $\epsilon_{pb}/\epsilon_{pp}$ and show that $\lambda_{c,a}$, $\lambda_{c,dc}$, and $\theta_0$ converge in the large $\epsilon_{bb}/\epsilon_{pp}$ limit. (See Fig.~\ref{fig:boundarystiffness} (d)-(f).) We find that $\lambda_c(\theta)<0$ (for all $\theta$) at large $\epsilon_{bb}/\epsilon_{pp}$ for all $N = 4$ configurations with FLW, since $\lambda_{c,dc} <0$ and $\lambda_{c,a} < |\lambda_{c,dc}|$. These results emphasize that cells with FLW become similar to cells with PBC (with $\lambda_c(\theta) <0$) in the large $\epsilon_{bb}$-limit. 
Thus, particle-wall interactions are essential for $\lambda_c(\theta)>0$.

\subsection{Shear modulus versus pressure for tessellated granular metamaterials}
\label{subsec:gp_cgm}

\begin{figure*}
    \centering
    \includegraphics[width=\linewidth]{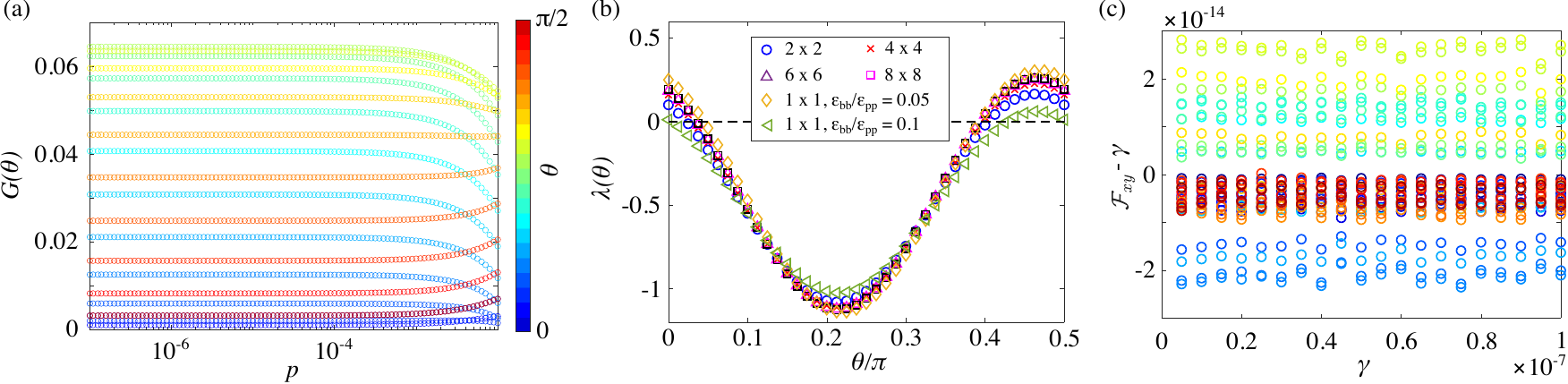}
\caption{(a) Shear modulus $G(\theta)$ versus pressure $p$ for a tessellated granular metamaterial composed of $\mathcal{N}_c=36$ cells containing the same $N=4$ disk configuration (i.e. configuration $5$ in Fig.~\ref{fig:fourparticle_config} (c)) with $\epsilon_{pb}/\epsilon_{pp} = 1$ and $\epsilon_{bb}/\epsilon_{pp} = 0.1$. Different colors correspond to the angle $\theta$ of the applied simple shear strain, which varies from $0$ (blue) to $\pi/2$ (red). (b) $\lambda(\theta)$ for tessellated granular metamaterials with $\epsilon_{pb}/\epsilon_{pp} = 1$ and $\epsilon_{bb}/\epsilon_{pp} = 0.1$ and different values of $\mathcal{N}_c$ (blue circles: $4$, red crosses: $16$, purple upward triangles: $36$, and magenta squares: $64$) and a single cell with $\epsilon_{bb/\epsilon_{pp}}=0.05$ (yellow diamonds) and $0.1$ (green left triangles). All cells contain the same disk configuration $5$ in Fig.~\ref{fig:fourparticle_config} (c) with FLW. (c) The off-diagonal term of the fitted displacement matrix $F_{xy}-\gamma$ in Eq.~\ref{eq:d2min} as a function of the applied simple shear strain $\gamma$ at $p = 10^{-7}$ for each cell (indicated by different colors) for the tessellated granular metamaterial in (a).}
    \label{fig:truss_lockin}
\end{figure*}

\subsubsection{Lock-in of single cell behavior}
\label{subsubsec:lock-in}

\begin{figure*}
    \centering
    \includegraphics[width=\linewidth]{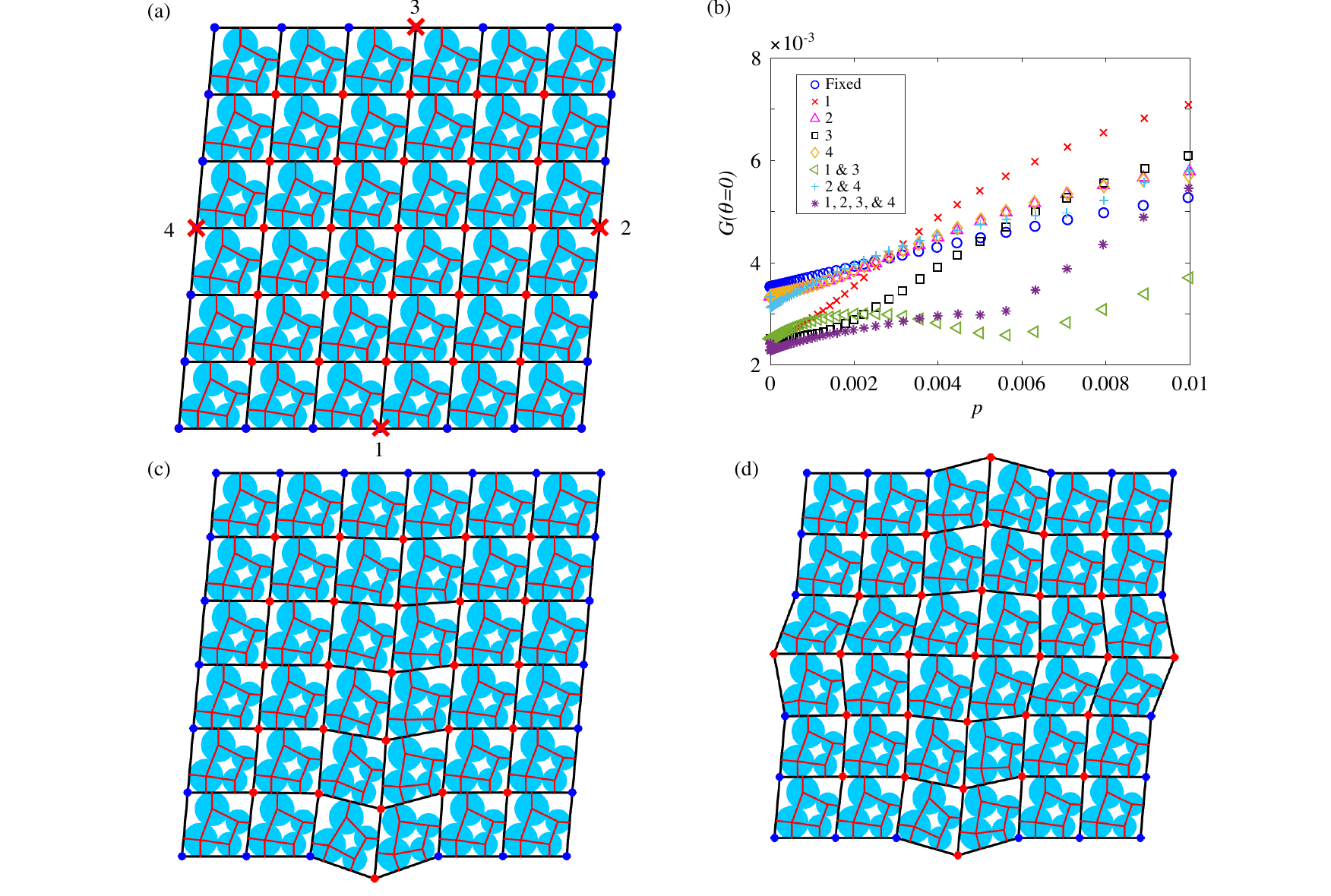}
\caption{(a) Tessellated granular metamaterial with $\mathcal{N}_c=36$ described in Fig.~\ref{fig:truss_lockin}, except now some of the wall endpoints marked by red crosses are no longer fixed after the applied isotropic compression and simple shear strain. (b) Shear modulus $G(0)$ measured at $\theta = 0$ as a function of pressure $p$ when different wall endpoints on the outer boundary are switched from fixed to mobile (blue circles: all outer wall endpoints are fixed, red crosses: endpoint $1$ is mobile, magenta upper triangles: endpoint $2$ is mobile, black squares: endpoint $3$ is mobile, yellow diamonds: endpoint $4$ is mobile, green left triangles: endpoints $1$ and $3$ are mobile, cyan crosses: endpoints $2$ and $4$ are mobile, and purple asterisks: endpoints $1$-$4$ are mobile). (c)-(d) Tessellated granular metamaterials at $p = 0.01$ when (c) endpoint $1$ is mobile, and (e) endpoints $1$-$4$ are mobile.}
    \label{fig:truss_loose}
\end{figure*}

We now study the pressure-dependence of the shear modulus $G(\theta)$ for tessellated granular metamaterials (Fig.~\ref{fig:trussexample}) constructed from multiple cells with flexible walls, $\epsilon_{pb}/\epsilon_{pp}=1$, and $\epsilon_{bb}/\epsilon_{pp}=0.1$. In Fig.~\ref{fig:truss_lockin} (a), we show $G(\theta)$ versus $p$ for $\mathcal{N}_c=36$ cells that each contain configuration $5$ from Fig.~\ref{fig:fourparticle_config} (c). Similar to the results for $G_c(\theta)$ for single cells, the mechanical response of tessellated granular metamaterials shows strong shear angle dependence. In particular, for some values of $\theta$, the slope of $G(\theta)$ versus $p$, $\lambda(\theta) >0$, and for other values, $\lambda(\theta) <0$. In Fig.~\ref{fig:truss_lockin} (b), we show that $\lambda(\theta)$ possesses weak system-size dependence as $\mathcal{N}_c$ is increased. $\lambda(\theta)$ for the multi-cell system in the large-$\mathcal{N}_c$ limit converges to $\lambda_c(\theta)$ for a single cell with FLW with $\epsilon_{bb}/\epsilon_{pp}=0.05$, which is half of the value for the multi-cell system. This result can be explained because each wall in the tessellated granular metamaterial is shared by its neighboring cell except for those on the exterior. Since $\lambda(\theta)$ for the tessellated granular metamaterial mimics that for single cells, these results indicate that we can lock-in the behavior of the shear modulus versus pressure for a single cell in tessellated granular metamaterials in large-$\mathcal{N}_c$ limit. In particular, we find finite regions of $\theta$ where $\lambda(\theta) >0$ in the large-$\mathcal{N}_c$ limit without particle rearrangements. The similarity of the mechanical response between the multi- and single-cell systems is caused by the fact that all of the single cells display similar displacements during applied simple shear strains. In Fig.~\ref{fig:truss_lockin} (c), we show that the non-affine displacements of each cell, caused by energy minimization after each applied simple shear strain, is negligible. The non-affine motion is obtained by measuring the difference in the off-diagonal element of the strain tensor of the interior vertices and the expected value from simple shear, $\mathcal{F}_{xy} - \gamma$, for each cell~\cite{falk98pre}. Previous studies have shown that non-affine particle motion strongly affects the magnitude of the shear modulus~\cite{ellenbroek09epl, jamming:WangPRE2021, staddon23softmatter}. Thus, near zero non-affine displacements of the cells indicates that the strain of the tessellated metamatarials has locked-in the strain of each cell. In general, the results for the tessellated granular metamaterials are qualitatively the same for any jammed packings that is used to construct the tessellated granular metmaterial.

\subsubsection{Constraints on outer wall vertices}
\label{subsubsec:truss_loose}

We also investigate how many constraints on the outer wall endpoints are necessary to enforce the mechanical response of the single cells in tessellated granular metamaterials. To address this question, we allow different wall endpoints on the outer boundary to relax during energy minimization following the applied isotropic compression and simple shear strain (see Fig.~\ref{fig:truss_loose} (a)). We find that $G(\theta)$ versus $p$ for tessellated granular metamaterials with even a single mobile outer wall endpoint deviates from that of the tessellated granular metamaterial where all outer wall endpoints are fixed. We show $G(0)$ versus $p$ as an example in  Fig.~\ref{fig:truss_loose} (b). Allowing the outer wall endpoints to move gives rise to buckling of tessellated granular metamaterials caused by compression, as shown in Fig.~\ref{fig:truss_loose} (c) and (d). This buckling induces changes in the shape of single cells compared to those of single cells with fixed outer wall endpoints during compression and shear, which causes the deviations in $G(\theta)$ versus $p$. Therefore, to ensure that tessellated granular metamaterials lock-in single-cell behavior, it is necessary to constrain all of the outer wall endpoints.

\begin{table*}
\renewcommand{\arraystretch}{1.5}
\caption{\label{table:summary}Summary of the results for the number of contacts $N_c$ and derivative of the shear modulus with respect to pressure, $\lambda_c$, of single cells for all three types of boundary conditions: PBC, FXW, and FLW.}
\begin{tabularx}{\textwidth} { 
   >{\centering\arraybackslash}X 
  | >{\centering\arraybackslash}X 
  | >{\centering\arraybackslash}X
  | >{\centering\arraybackslash}X }
 \hline \hline
 & PBC & FXW & FLW \\
\hline
Number of contacts at jamming onset, $N_c$ & Isotropic or hyperstatic & Isotropic or hyperstaitc & Isotropic, hyperstatic, or hypostatic \\
Sign of $\lambda_c$ & $\lambda_c < 0$ for $N_c > 4$ & $\lambda_c < 0$ or $\lambda_c > 0$ & $\lambda_c < 0$ or $\lambda_c > 0$ \\
\hline \hline
\end{tabularx}
\end{table*}

\section{Conclusions and future directions}
\label{sec:conclusions}

In the large-system limit, the shear modulus $G$ of static packings of spherical particles increases with pressure due to frequent particle rearrangements and non-affine particle motion that enable the packings to increase their contact number with increasing pressure.  In this work, we investigate a novel class of granular materials, tessellated granular metamaterials, that allow us to control the slope of the shear modulus versus pressure by preventing particle rearrangements even in the large-system limit.  We focus on tessellated granular metamaterials in two dimensions, which are collections of $\mathcal{N}_c$ coupled cells that each contain $N$ bidisperse disks enclosed by four physical walls. In particular, we can design tessellated granular metamaterials with negative slope of the shear modulus with pressure even in the large-$\mathcal{N}_c$ limit. 

We first studied the mechanical properties of single cells with three sets of boundary conditions: PBC, FXW, and FLW. Packings with small $N$ do not undergo frequent particle rearrangements, and thus we enumerated all possible mechanically stable cells with all three boundary conditions for $N \leq 8$. We find that the mechanically stable cells with PBC and FXW are either isostatic or hyperstatic, while those with FLW can be hypostatic, as well as isostatic and hyperstatic. The hypostatic cells with FLW are stabilized by quartic modes, as found for hypostatic packings of non-spherical and deformable particles. Second, we showed that the angle-dependent shear modulus of single cells depends linearly on pressure, $G_c(\theta) = \lambda_c(\theta) p + G_{c0}(\theta)$. Further, the slope of the shear modulus versus pressure for single cells is strongly anisotropic, i.e. $\lambda_c(\theta) = \lambda_{c,dc} + \lambda_{c,a} \sin(4(\theta - \theta_0))$ and $\lambda_{c,a} \sim \lambda_{c,dc}$. We find that $\lambda_c(\theta)<0$ for single cells in PBC with $N > 4$. In contrast, cells with FXW and FLW and small $N$ can possess either $\lambda_c(\theta) >0$ or $\lambda_c(\theta) <0$.  However, the probability of obtaining cells with $\lambda_c(\theta) >0$ vanishes in the large-$N$ limit. These results are summarized in Table.~\ref{table:summary}. After studying the mechanical response of single cells, we investigated the shear modulus of tessellated granular metamaterials formed by connecting many single cells with flexible walls. We showed that we can lock-in the mechanical response of single cells in tessellated granular metamaterials. The ability to lock-in the mechanical response of single cells in multi-cell systems is reduced if the outer wall endpoints are free to move during energy minimization after applied deformations. These results demonstrate that we can build large-scale granular metamaterials whose mechanical properties do not change after repeated cycles of compression and decompression, as well as positive and negative simple shear strain, since particle rearrangements are eliminated. 

These findings raise many interesting directions for future research. First, we found that the mechanical response of both single and multiple-cell granular systems is highly anisotropic. To fully understand the pressure-dependent mechanical properties of anisotropic materials in two dimensions, we must characterize all $6$ stiffness matrix components as a function of pressure, which requires additional mechanical tests in addition to simple shear as a function of $\theta$. Doing so will also aid in the development of a continuum elasticity theory for tessellated granular materials. Second, for the current studies, we fixed all of the outer wall endpoints during energy minimization to enforce nearly affine simple shear of tessellated granular metamaterials.  However, when the outer wall endpoints are not fixed, the individual cells can change their shape during energy minimization that follows the applied compression and simple shear strain. Thus, it will be interesting to study and predict the pressure dependence of $G_c(\theta)$ of single cells when the outer wall endpoints are free to move or bending energy is included between adjacent endpoints to generate cells with arbitrary shapes. Third, we have focused on tessellated granular metamaterials composed of identical single cells.  In future studies, we will consider tessellated granular metamaterials composed of single cells with different disk configurations and boundaries with varied $\epsilon_{pb}$ and $\epsilon_{bb}$ to understand how the mechanical properties of single cells determine the mechanical properties of the multi-cell system. Fourth, we have only considered square tessellated metamaterials. We can also investigate tessellated metamaterials with different numbers of rows and columns of cells, which can provide more tunability of the pressure-dependent mechanical response. Finally, we will extend our studies of tessellated granular metamaterials to three dimensions. In three dimensions, there are three principal simple shear directions, instead of one in two dimensions, which provides additional ways to design tessellated granular metamaterials. For example, we can create strongly anisotropic tessellated granular metamaterials by having some cells possess $\lambda_c(\theta) >0$ in one shear direction, others possess $\lambda_c(\theta) <0$ in another shear direction, and others possess $\lambda_c(\theta) >0$ in the remaining shear direction. 

\section*{Acknowledgments}

The authors acknowledge support from ONR Grant No. N00014-22-1-2652 and NSF Grant No. DMREF-2118988. This work was also supported by the High Performance Computing facilities operated by Yale’s Center for Research Computing.

\appendix

\section{Isostaticity \& quartic modes in a single cell}
\label{sec:isostaticity}

In this Appendix, we discuss the number of degrees of freedom $N_d$ in cells for all three types of boundary conditions and quartic modes in hypostatic disk packings with FLW. 

\begin{figure}
    \centering
    \includegraphics[width=\linewidth]{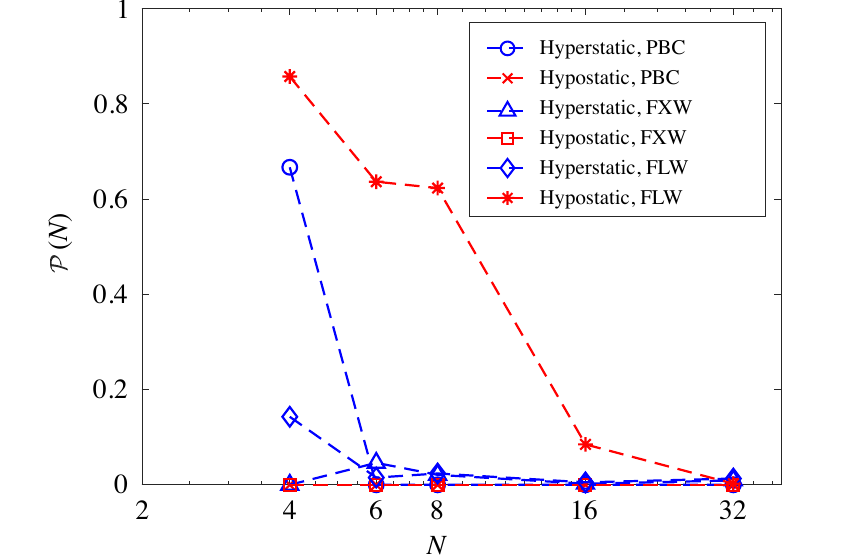}
\caption{Probability ${\cal P}(N)$ (over $10^4$ single cells) of obtaining a given number of contacts, hyperstatic ($N_c > N_d$) or hypostatic ($N_c < N_d$), for single cells with PBC, FXW, or FLW as a function of system size $N$.}
    \label{fig:hyperhypoprob}
\end{figure}

In PBC, $N_{d} = 2N' - 2 + 1 = 2N' - 1$, where $N' = N - N_r$ and $N_r$ is the number of rattler disks. In this expression, the $-2$ comes from the two global translational degrees of freedom in periodic boundary conditions; the $+1$ comes from the size degree of freedom of the disks while maintaining fixed diameter ratio. The $+1$ can also be interpreted as a state of self-stress~\cite{pellegrino86stability}. For the degrees of freedom in cells with FXW and FLW, we must include the degrees of freedom of the wall endpoints, as well as the disks: $N_{d} = 2N' + 2N_v - N_B - 3 + 1$. Here, $N_v = 4$ is the number of wall endpoints, $N_B$ is the number of constraints associated with the walls, the $-3$ comes from the two rigid-body translational and one rotational degree of freedom, and the $+1$ again comes from the size degree of freedom of the disks. For FLW, four springs connect the wall endpoints, and hence $N_B = 4$. For FXW, in addition to the length constraint for each wall, the angle between any two neighboring walls is also fixed, and thus $N_B = 5$. Hence, $N_{d} = 2N' + 1$ for cells with FXW and $N_{d} = 2N' + 2$ for FLW. For isostatic packings, the total number of contacts satisfies $N_c = N_d$. A packing is hyperstatic when $N_c > N_d$ and hypostatic when $N_c < N_d$. We show that the probability of obtaining a hyperstatic or hypostatic cell decreases with increasing $N$ for all three boundary conditions as shown in Fig.~\ref{fig:hyperhypoprob}. We note that there are a finite number of hypostatic packings in cells with FLW even at $N \le 16$, which highlights the effect of soft physical walls on the structural and mechanical properties of jammed granular materials.

\begin{figure}
    \centering
    \includegraphics[width=\linewidth]{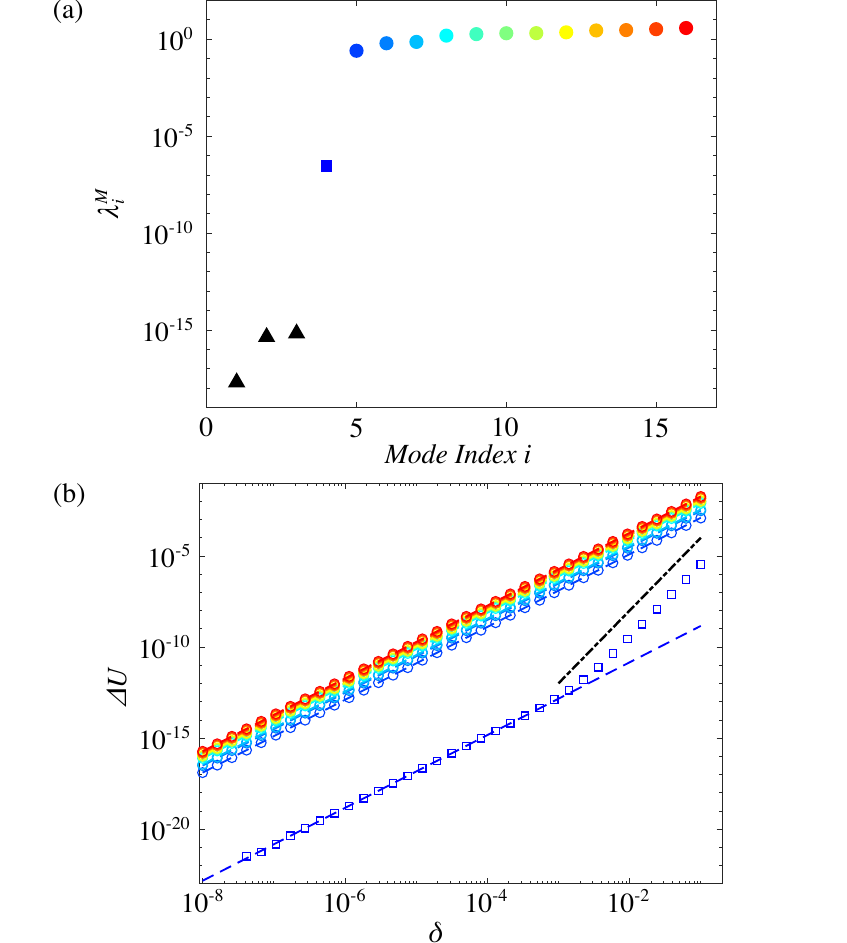}
    \caption{(a) Sorted eigenvalues (from smallest to largest) of the dynamical matrix $\lambda^M$ for configuration 5 with FLW boundary conditions in Fig. 3(c). The first three eigenvalues (corresponding to two rigid-body translations and one rigid-body rotation) are less than $10^{-15}$ and decrease in magnitude as we improve the force balance condition. The fourth eigenvalue ($i=4$) corresponds to a ``quartic mode'' and those with $i>4$ correspond to quadratic modes. (b) Change in potential energy $\Delta U$ in response to perturbations along eigenvectors ${\hat{\lambda}}^M$ plotted as a function of the perturbation amplitude $\delta$. The dashed and dot-dashed lines have slope $2$ and $4$, respectively. The colors of the symbols in (b) correspond to perturbations along the eigenvectors associated with eigenvalues with the same colored symbols in (a).}
    \label{fig:quarticmodes}
\end{figure}

``Quartic'' modes stabilize hypostatic jammed packings.  They can be illustrated through the expansion of the total potential energy of a jammed disk packing $U$ in terms of small displacements around its local energy minimum, $U({\vec{R}}_0)$, where ${\vec{R}}_0$ denotes the equilibrium positions of all particles and wall vertices. If we perturb the packing to a new set of positions, $\vec{R}={\vec{R}}_0+\delta\hat{u}$, where $\delta$ is the amplitude of the perturbation and $\hat{u}$ is a unit vector characterizing the perturbation direction, $U$ can be approximated by a Taylor expansion:
\begin{equation}
\begin{split}
U(\vec{R}) = & U({\vec{R}}_0) + \delta \sum_{i}{\frac{\partial U}{\partial R_i}u_i} +\frac{1}{2} \delta^2 \sum_{i,j}{\frac{\partial^2U}{\partial R_i\partial R_j}u_i u_j} \\
& + \frac{1}{6} \delta^3 \sum_{i,j,k}{\frac{\partial^2U}{\partial R_i\partial R_j\partial R_k}u_i u_j u_k} \\
& + \frac{1}{24} \delta^4 \sum_{i,j,k,l}{\frac{\partial^2U}{\partial R_i\partial R_j\partial R_k\partial R_l}u_i u_j u_k u_l} + \cdots,
\end{split}
\end{equation}
where the derivatives are evaluated at ${\vec{R}}_0$. The second term on the right-hand side (RHS) of this equation is zero since the jammed packing is in force balance. The third term on the RHS includes the dynamical matrix $M = \frac{\partial^2U}{\partial R_i\partial R_j}$, whose eigenvalues $\lambda^M$ give the energy associated with perturbations along the associated eigenvectors ${\hat{\lambda}}^M$. In isostatic and hyperstatic packings with $N_c \ge N_d$, all non-trivial eigenvalues $\lambda^M$ are non-zero and $\Delta U = U({\vec{R}}) - U({\vec{R}}_0)$ increases quadratically with amplitude $\delta$ for perturbations along ${\hat{\lambda}}^M$: $\Delta U = \frac{1}{2} \lambda^M \delta^2$. In hypostatic packings, some of the eigenvalues of $M$ are much smaller than the non-trivial quadratic eigenvalues, yet they are larger than the trivial eigenvalues corresponding to rigid-body translations and rotations. (For example, see the fourth eigenvalue of $M$ in Fig.~\ref{fig:quarticmodes}(a).) The number of these modes matches the number of missing contacts, $N_d-N_c$.  When perturbing along eigenvectors ${\hat{\lambda}}^M$ associated with one of these small non-trivial eigenvalues, $\Delta U$ first scales quadratically with $\delta$ and then scales quartically with $\delta$ at larger values of $\delta$ as shown in Fig.~\ref{fig:quarticmodes}(b). In previous studies~\cite{schreck12pre}, we have shown that the characteristic $\delta_c$ at which the scaling of $\Delta U$ changes from quadratic to quartic decreases with pressure, and thus at zero pressure, $\Delta U$ scales quartically with $\delta$. (These previous results emphasize that the fourth term in the Taylor expansion containing third derivatives of $U$ is small.)  Since these modes scale quartically with $\delta$ at zero pressure, we label them as ``quartic modes.''

\section{Stiffness matrix after rotation}
\label{sec:C_mat_rot}

In this Appendix, we show the angular dependence of all elements of the stiffness matrix ${\hat C}$, which relates stress and strain. At a predefined orientation with $\theta = 0$, we can calculate the stiffness matrix ${\hat C}(0)$.  After rotating the configuration by an angle $\theta$ clockwise, the stiffness matrix becomes ${\hat C}(\theta)= {\hat R}^T(\theta){\hat C}(0){\hat R}(\theta)$, where
\begin{equation}
    \label{eq:rc_matrix}
    {\hat R}(\theta) = \begin{pmatrix}
    	\cos^2\theta & \sin^2\theta & -\frac{1}{2}\sin2\theta \\
		\sin^2\theta & \cos^2\theta & \frac{1}{2}\sin2\theta \\
		\sin2\theta & -\sin2\theta & \cos2\theta
        \end{pmatrix}.
\end{equation}
Using Eq.~\ref{eq:rc_matrix}, we find the following angle-dependent stiffness matrix elements : 
\begin{equation}
    \label{eq:c11_rot}
    \begin{split}
        \hat{C}_{11}(\theta) = & \hat{C}_{11}(0) \cos^4\theta + \hat{C}_{22}(0) \sin^4\theta \\
                               & + \hat{C}_{33}(0) \sin^2(2\theta) + \frac{1}{2} \hat{C}_{12}(0) \sin^2(2\theta) \\
                        & + 2\hat{C}_{13}(0) \sin(2\theta) \cos^2\theta \\
                        & + 2\hat{C}_{23}(0) \sin(2\theta) \sin^2\theta,
    \end{split}
\end{equation}
\begin{equation}
    \label{eq:c12_rot}
    \begin{split}
	\hat{C}_{12}(\theta) = & \left( \frac{1}{4} \left( \hat{C}_{11}(0) + \hat{C}_{22}(0) \right) - \hat{C}_{33}(0) \right) \sin^2(2\theta) \\
					  & + \hat{C}_{12}(0) \left( \sin^4\theta + \cos^4\theta \right) \\
					  & + \frac{1}{2} \left( \hat{C}_{23}(0) - \hat{C}_{13}(0) \right) \sin(4\theta),
    \end{split}
\end{equation}
\begin{equation}
    \label{eq:c13_rot}
    \begin{split}
	\hat{C}_{13}(\theta) = & -\frac{1}{2} \hat{C}_{11}(0) \sin(2\theta) \cos^2\theta + \frac{1}{2} \hat{C}_{22}(0) \sin(2\theta) \sin^2\theta \\
					  & + \frac{1}{2} \hat{C}_{33}(0) \sin(4\theta) + \frac{1}{4} \hat{C}_{12}(0) \sin(4\theta) \\
					  & + \hat{C}_{13}(0) \cos^2\theta \left( 2\cos(2\theta) - 1 \right) \\
					  & + \hat{C}_{23}(0) \sin^2\theta \left( 2\cos(2\theta) + 1 \right),
    \end{split}
\end{equation}
\begin{equation}
    \label{eq:c22_rot}
    \begin{split}
	\hat{C}_{22}(\theta) = & \hat{C}_{11}(0) \sin^4\theta + \hat{C}_{22}(0) \cos^4\theta + \hat{C}_{33}(0)\sin^2(2\theta) \\
	                  & + \frac{1}{2} \hat{C}_{12}(0) \sin^2(2\theta) - 2\hat{C}_{13}(0) \sin(2\theta) \sin^2\theta \\
                        & - 2\hat{C}_{23}(0) \sin(2\theta) \cos^2\theta,
    \end{split}
\end{equation}
\begin{equation}
    \label{eq:c23_rot}
    \begin{split}
	\hat{C}_{23}(\theta) = & -\frac{1}{2} \hat{C}_{11}(0) \sin(2\theta) \sin^2\theta + \frac{1}{2} \hat{C}_{22}(0) \sin(2\theta) \cos^2\theta \\
					  & - \frac{1}{2} \hat{C}_{33}(0) \sin(4\theta) - \frac{1}{4} \hat{C}_{12}(0) \sin(4\theta) \\
					  & + \hat{C}_{13}(0) \sin^2\theta \left( 2\cos(2\theta) + 1 \right) \\
					  & + \hat{C}_{23}(0) \cos^2\theta \left( 2\cos(2\theta) - 1 \right),
	\end{split}
\end{equation}

\begin{equation}
    \label{eq:c33_rot}
    \begin{split}
	\hat{C}_{33}(\theta) = & \frac{1}{4} \left( \hat{C}_{11}(0) + \hat{C}_{22}(0) - 2\hat{C}_{12}(0) \right) \sin^2 \left( 2\theta \right) \\
					  & + \frac{1}{2} \left( \hat{C}_{23}(0) - \hat{C}_{13}(0) \right) \sin\left( 4\theta \right) \\
					  & + \hat{C}_{33}(0) \cos^2\left( 2\theta \right).
    \end{split}
\end{equation}
Eqs.~\ref{eq:c11_rot}-\ref{eq:c33_rot} show that generally all six elements of the reference stiffness matrix contribute to each ${\hat C}$ element at a given angle $\theta$. Therefore, in anisotropic materials, it is important to track all ${\hat C}$ elements to fully characterize their mechanical properties.

\section{Relation between shear modulus and mixed shear strain derivatives}
\label{sec:twoparticle}

\begin{figure}
    \centering
    \includegraphics[width=\linewidth]{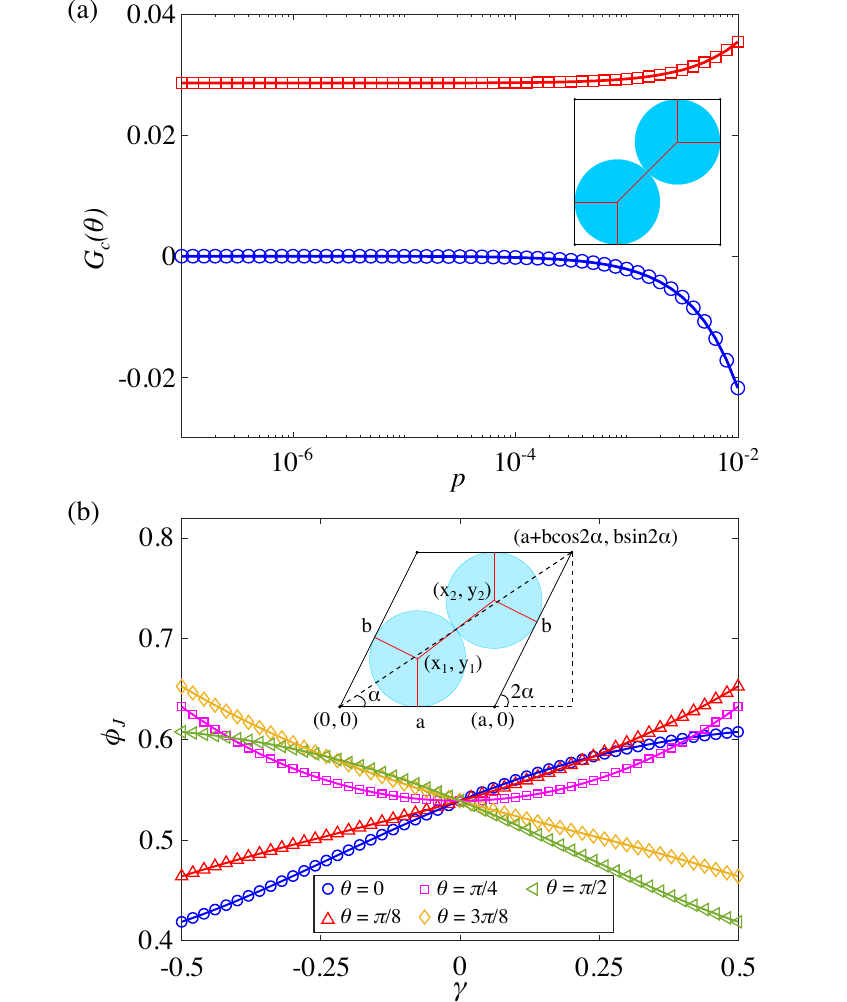}
\caption{(a) Shear modulus $G_c(\theta)$ for a single cell containing a monodisperse $N = 2$ disk packing with FXW at shear angle $\theta = 0$ (red squares and red solid line) and $\pi/4$ (blue circles and blue solid line) as a function of pressure $p$. The squares and circles show $G_c(\theta)$ calculated using $G_c = d\Sigma_c/d\gamma$, while the solid lines show $G_c(\theta)$ calculated using Eq.~\ref{eq:G_phi_gamma}.  The inset shows the $N = 2$ cell at simple shear strain $\gamma = 0$ and $\theta = 0$. (b) Packing fraction $\phi_J$ of the single cell in (a) at jamming onset as a function of $\gamma$ at several values of $\theta$ as indicated by the different colors and symbols. The symbols and solid lines correspond to the results from the numerical simulations and analytical calculations using Eqs.~\ref{eq:length_a}-\ref{eq:phi_gamma_theta}, respectively. The inset provides an illustration of a monodisperse $N=2$ packing in a cell with FXW boundary conditions at jamming onset. The disk centers are specified by $(x_1,y_1)$ and $(x_2,y_2)$ and the lengths of the parallelogram side walls are $a$ and $b$ with angle $2 \alpha$ between them.}
    \label{fig:twoparticle}
\end{figure}

In this Appendix, we verify that the pressure-dependence of the single-cell shear modulus is related to the variation of the packing fraction at jamming onset $\phi_J$ with simple shear strain $\gamma$ as shown in previous studies~\cite{zhang21pre}. We illustrate this relationship using a single cell containing the $N=2$ monodisperse disk packing with FXW in the inset to Fig.~\ref{fig:twoparticle} (a) since $\phi_J(\gamma)$ can be calculated analytically for this case. The shear modulus can be written in terms of three mixed derivatives of the simple shear strain: 
\begin{equation}
    \label{eq:G_phi_gamma}
    \begin{split}
    G & = \left( \frac{d\Sigma}{d\gamma} \right)_{\phi} = \frac{1}{A} \left( \frac{ d\left( \frac{dU}{d\gamma} \right)_{p} }{d\gamma} \right)_{\phi} \\
      & - \frac{p}{\phi} \left( \frac{ d\left( \frac{d\phi}{d\gamma} \right)_{p} }{d\gamma} \right)_{\phi} - \frac{1}{\phi} \left( \frac{dp}{d\gamma} \right)_{\phi} \left( \frac{d\phi}{d\gamma} \right)_{p}.
    \end{split}
\end{equation}
In Fig.~\ref{fig:twoparticle} (a), we demonstrate that Eq.~\ref{eq:G_phi_gamma} still holds for single cells with FXW.

The second term in Eq.~\ref{eq:G_phi_gamma} is proportional to $p$, with $\left( d \left( \frac{d\phi}{d\gamma} \right)_{p} / d\gamma \right)_{\phi} > 0$ for single cells with $N \ge 6$ and PBC~\cite{jamming:VanderWerfPRL2020}. In contrast, the first and third terms in Eq.~\ref{eq:G_phi_gamma} do not possess strong $p$-dependence. Therefore, the second derivative of $\phi_J(\gamma)$ typically determines whether $G$ will increase or decrease with $p$. In particular, if $\phi_J(\gamma)$ is concave upward, $G$ decreases with increasing $p$, and vice versa.

For single cells containing $N=2$ monodisperse disks at jamming onset with FXW boundary conditions, we can analytically determine the packing fraction at jamming onset $\phi_J$ as a function of the simple shear strain $\gamma$ and shear angle $\theta$. Consider a square box (with initial side lengths $L_0=1$) whose vertices are located at $(0, 0)$,  $(1, 0)$, $(1, 1)$, and $(0, 1)$. After rotation by $\theta$ clockwise about the origin, these four points transform into $(0, 0)$, $(\cos{\theta}, \sin{\theta})$, $(\cos{\theta}+\sin{\theta}, -\sin{\theta}+\cos{\theta})$, and $(\sin{\theta}, \cos{\theta})$. After we apply a simple shear strain $\gamma$ to the rotated box, the four points become $(0, 0)$, $(\cos{\theta}-\gamma \sin{\theta}, -\sin{\theta})$, $(\cos{\theta}+\sin{\theta}+\gamma(-\sin{\theta}+\cos{\theta}), -\sin{\theta}+\cos{\theta})$, and $(\sin{\theta}+\gamma\cos{\theta}, \cos{\theta})$. The resulting box is a parallelogram with unit area and side lengths,
\begin{equation}
    \label{eq:length_a}
    a = \sqrt{1 - 2\gamma \sin\theta \cos\theta + \gamma^2\sin^2\theta},
\end{equation}
\begin{equation}
    \label{eq:length_b}
    b = \sqrt{1 + 2\gamma \sin\theta \cos\theta + \gamma^2\cos^2\theta}.
\end{equation}
To obtain $\phi_J(\theta, \gamma)$, we need to find the radius of the circle $r$ such that the two disks fit within the parallelogram and are mechanically stable. (See Fig.~\ref{fig:twoparticle}(b) inset, where we have rotated the parallelogram by $\theta$ counterclockwise to simplify the calculation.) The centers of the two disks are located at $(x_1, y_1)$ and $(x_2, y_2)$, where $x_1 = r\cot{\alpha}$, $y_1 = r$, $x_2 = a + b\cos{2\alpha} - r\cot{\alpha}$, and $y_2 = b\sin{2\alpha} - r\cot{\alpha}$, with
\begin{equation}
    \label{eq:alpha}
    \alpha = \frac{1}{2} \cos^{-1} \left( \frac{\gamma (\cos^2\theta - \sin^2\theta) - \gamma^2\sin\theta\cos\theta}{ab} \right).
\end{equation}
The distance between the two disks must equal $2r$, and thus ${(x_1-x_2)}^2+{(y_1-y_2)}^2={(2r)}^2$. This expression yields a quadratic equation for $r$: $Ar^2+Br+C=0$, where $A$, $B$, and $C$ are given by
\begin{equation}
    \label{eq:A}
    A = 4\cot^2\alpha,
\end{equation}
\begin{equation}
    \label{eq:B}
    B = -4 (a \cot\alpha + b (\cot\alpha \cos(2\alpha) + \sin(2\alpha))),
\end{equation}
and
\begin{equation}
    \label{eq:C}
    C = a^2 + 2 a b \cos(2\alpha) + b^2.
\end{equation}
The two solutions for $r$ are
\begin{equation}
    \label{eq:r1}
    r_{1,2} = \frac{-B\pm\sqrt{B^2-4AC}}{2A}.
\end{equation}
The first solution results in circles that are outside of the parallelogram.  Thus, $r=r_2$ and
\begin{equation}
    \label{eq:phi_gamma_theta}
    \phi_J(\theta,\gamma) = \pi \frac{\left(-B - \sqrt{B^2 - 4AC} \right)^2}{2 A^2}.
\end{equation}
We verify in Fig.~\ref{fig:twoparticle} (b) that $\phi_J(\theta,\gamma)$ determined by the numerical simulations matches that predicted by Eq.~\ref{eq:phi_gamma_theta}. At $\gamma = 0$, which is where $G_c(\theta)$ is measured throughout the main text, we find that the $\phi_J(\gamma)$ is concave downward at $\theta = 0$ and concave upward at $\theta = \pi/4$ (Fig.~\ref{fig:twoparticle} (b)). Thus, since $\lambda_c(\theta)$ switches sign, we expect a saddle point to occur in the $\phi_J(\gamma,\theta)$ plane between $\theta =0$ and $\pi/4$. In future studies, we will apply a similar approach in Eq.~\ref{eq:G_phi_gamma} to obtain the pressure-dependence of all elements of the stiffness matrix.

\bibliography{main}

\end{document}